\documentclass[a4paper]{article}
\usepackage[11pt]{extsizes}
\usepackage{fullpage} 
\usepackage{pstricks}
\usepackage{multicol}
\usepackage{graphicx}
\usepackage{hyperref}
\usepackage[version=3]{mhchem}
\usepackage{amsmath}
\usepackage{amssymb} 
\usepackage{amsfonts}
\usepackage{epstopdf}
\usepackage{color}
\usepackage{authblk}
\usepackage[margin=1.5cm,top=2.5cm,bottom=2.8cm,centering]{geometry}

\title{\begin{flushleft}\textbf{Reductionist approach to chemical rate constants using conditional energy probabilities\footnote{Reference: Michel D. 2019. Reductionist approach to chemical rate constants using conditional energy probabilities. European Journal of Physics 40, 055103. \href{https://doi.org/10.1088/1361-6404/ab272e}{DOI}.}}\end{flushleft}}
\date{}
\author{}
\begin{document}
\maketitle
\vspace*{-1.5cm}
\noindent
Denis Michel\\

\noindent
\small Univ Rennes, Inserm, EHESP, Irset UMR 1085, Rennes, France. denis.michel@live.fr.

\vspace*{0.5cm}
\noindent

\begin{multicols}{2}
\noindent
\textbf{Different rate theories yielded similar forms of rates constants consistent with the phenomenological Arrhenius law, although they were derived from various branches of physics including classical thermodynamics, statistical and quantum mechanics. This convergence supports the validity of the Arrhenius law but also suggests the existence of an even simpler underlying principle. A reductionnist approach is proposed here in which the energetic exponential factor is a conditional probability of sufficient energy and the pre-exponential factor is the frequency of recurrence of the configuration favorable to the reaction, itself proportional to a configurational probability in a chaotic system. This minimalist while rigorous mathematical approach makes it possible to bypass certain questionable postulates of more sophisticated theories and clarifies the meaning of the different types of energies used: activation energy, threshold energy and chemical energy.}

\section{Introduction}
Certain physical laws have been predicted theoretically before being verified experimentally. But historically, this is not the case of the most widely used law in chemical kinetics: the Arrhenius constant, which has been laboriously constructed to best fit to chemical transformations \cite{Laidler1984}. The subject of this paper is precisely to imagine how the current form of rate constants might have been predicted before undertaking experimental tests. Such a theoretical exercise is of course easier when the expected result is already know, but it will nevertheless prove of pedagogical interest and useful for reconsidering rate theories. The classical form of the Arrhenius law was published in 1889 \cite{Arrhenius} and anticipated by van't Hoff in 1884 \cite{van't Hoff}, but van't Hoff acknowledged that he did not enter into theoretical considerations \cite{van't Hoff}. This equation related to temperature through the Boltzmann factor, has proved very effective for modeling chemical reactions, but also simple experiments unrelated to chemistry such as a mechanical simulator allowing one to count the passages of a randomly propelled ping-pong ball between two different levels of a plateau \cite{Prentis}. Several attempts of formalization have been developed a posteriori, based on thermodynamics, statistical mechanics and quantum mechanics, including diffusion from an energy well \cite{Kramers}, reactive flux \cite{Chandler} and transition path \cite{Vanden-Eijnden} theories. These different approaches yielded similar formulas, which can of course be explained in part by the will of the authors to recover the efficient general form, but in addition, the same form is obtained for reactions of any order \cite{Tolman} and the same maximal quantum frequency can be derived as well by envisioning the reaction coordinate as a vibration or as a translation \cite{Eyring}. This convergence of approaches towards a single general law justifies the Arrhenius law and also suggests the existence of a founder principle reduced to its mathematical essence. Chemical reactions will be conceived as stochastic transitions whose frequency is just determined by the waiting time of two simultaneous favorable conditions, one configurationnal and the other energetic. In practice, reactants must not only be well positioned and oriented for the reaction, but also receive from their environment a sufficient amount of energy, higher than the average ambient energy, which occasionally happens by fluctuations. Remarkably, the simplest and in fact the only rigorous mathematical way of meeting by chance these two conditions, directly gives the law of Arrhenius, thus bypassing more complex theories and making unnecessary some postulates. In particular, this direct construction of the Arrhenius law does not use the chemical intermediate in equilibrium with the reactant called activated complex in the transition state theory (TST), whose existence somewhat breaks the notion of elementary reaction and mixes equilibrium and transient phenomena. The alternative presentation also helps clarifying the different chemical energy values used in rate theories, such as molecular energies and the transition state energy. A primary characteristic of the present proposal is to conform to the concept of unique stochastic chemical event, which is strangely ignored in diffusion, random walk and reaction path theories.

\section{Elementary chemical events}
Elementary reactions are successfully modeled as stochastic events characterized by a waiting time, but whose progress is itself instantaneous. The only mathematical law possible to describe the memoryless nature of elementary stochastic events is the exponential probability \cite{Michel2013}. Indeed, an event is memoryless if the fact that it did not take place for a time $ t_{1} $, does not influence its occurrence during the following time $ t_{2} $. This condition reads

\begin{subequations} 
\begin{equation} \forall \  t_{1},t_{2} \in \mathbb{R}_{+}, P(X \geqslant t_{1}+t_{2}\mid X \geqslant t_{1})=P(X \geqslant t_{2}) \end{equation} 
and since the left hand term is also
\begin{equation} \begin{split}
 P(X\geqslant t_1+t_2 \mid X\geqslant t_1)& =\dfrac{P({X\geqslant t_1+t_2} \cap {X\geqslant t_1} )}{P(X\geqslant t_1)}\\& = \dfrac{P(X\geqslant t_1+t_2)}{P(X \geqslant t_1)} \end{split} \end{equation}
the condition of lack of memory implies
\begin{equation} P(X \geqslant t_{1}+t_{2})=P(X\geqslant t_{1}) \ P(X \geqslant t_{2}) \end{equation}
The law satisfying this relationship is exponential, such that
\begin{equation} \large{\textup{e}}^{-k(t_{1}+t_{2})}=\large{\textup{e}}^{-kt_{1}} \  \large{\textup{e}}^{-kt_{2}} \end{equation}
\end{subequations}
\noindent
where $ k $ appears as a constant necessary to adimension the exponents, whose unit is in time$ ^{-1} $ and whose reciprocal is the expectation of this probability $ \left \langle t \right \rangle=1/k $. Under this mathematical assumption, the arrow traditionally used to denote a chemical transformation does not represent a path and the misleading term velocity sometimes used, should be replaced by mean frequency, which more clearly focuses on the intervals between the events, which are themselves instantaneous jumps without duration of their own. Therefore, any introduction of an intermediate compound in this scheme necessarily contradicts the mathematical model. This is the case when a catalyst in involved in the reaction. As discussed later, this is also the case of the celebrated activated complex understood as a molecular reaction intermediate. More generally, the reaction is not elementary in diffusion, Brownian motion and mean first passage theories, in which it is envisioned as a series of steps with a ratchet endpoint or a Markovian walk with an absorbing boundary. Such chains have non-exponential mean first arrival times. Serial microreactions, each one with its own energy barrier and gravitating or descending energy gradients, are likely to play essential roles in the accuracy of biochemical interactions \cite{Michel2016}, but the reciprocal of their mean time is not a rate constant but a combination of rate constants. Elementary reactions are in reality often embedded in complex reactional schemes and indistinguishable. In such situations, it may be preferable to abandon the Arrhenius equation and to use for instance the empirical three-parameter extended Arrhenius equation for gas kinetics, or in a case-by-case manner operational formulas derived from quasi-equilibrium theories \cite{Peters}. An elementary reaction with a single energy barrier may appear theoretical as it assumes the absence of reaction intermediates; but the intuitive view of such intermediates as conventional chemical compounds, would not be in agreement with quantum physics, since their lifetime would be less than the minimum time window $ \tau = \dfrac{h}{k_{B} T} $ (where $ k_{B} $ is the Boltzmann constant and $ h $ is the Planck constant). But before discussing the existing theories, let us describe \textit{de novo} the simple approach proposed here.

\section{The Arrhenius preexponential factor is based on the exponential distribution of events}

The Arrhenius formula
\begin{equation}  \label{Arrhenius}  k = A \ \large{\textup{e}}^{-\dfrac{E_{a}}{k_{B}T}} \end{equation}  

contains a preexponential factor $ A $ and an exponential factor including the activation energy ($ E_{a} $). The preexponential factor depends on the number of possible configurations of the reactant.
If a given reactant exists under $ \Omega $ spatial (not energetic) configurations, out of which only one allows the reaction to proceed, and if the different configurations of the reactants appear stochastically and are separated by a minimal time step $ \tau $, then the reactional configuration (event $ X $ of probability $ 1/\Omega $) reappears following a random (exponential) time distribution with an average waiting time $ \left \langle t \right \rangle $ such that

\begin{subequations}
\begin{equation}  \left \langle t \right \rangle = \Omega \tau \end{equation}
giving the preexponential factor
\begin{equation} A=\dfrac{1}{\left \langle t \right \rangle } \end{equation}
\end{subequations}

The general term $ \Omega $ must then be specified in each type of reaction by enumerating the molecular configurations involved. This exponential distribution of events rigorously reflects Boltzmann's vision of the memoryless microscopic chaos. The minimal waiting time for an elementary reaction is the universal time step $ \tau $ predicted by the so-called quantum uncertainty principle of Eisenberg. Indeed, the fundamental units of energy $ \varepsilon $  and time $ \tau $ are conjugated in Planck boxes of undeterminacy $ \varepsilon \tau = h $. Application of this rule to thermal physics gives $ \tau = h/(k_{B}T)$, regardless of the form of energy considered: \label{tau}

\begin{itemize}
\item For a resonator of frequency $ \nu $, when the thermal and quantum energies equalize, $ k_{B}T = h\nu $ \\ and we have $ \tau= \dfrac{1}{\nu} = \dfrac{h}{k_{B}T} $
\item For translational energy, the wavelength of de Broglie $ \lambda = \dfrac{h}{\sqrt{2\pi m k_{B}T}} $ is covered at the velocity predicted by kinetic theory $ v  = \sqrt{\dfrac{k_{B}T}{2 \pi m}} $,\\ in a time of $ \tau= \dfrac{\lambda}{v} = \dfrac{h}{k_{B}T} $. 
\end{itemize}

This maximum frequency can be found in several ways \cite{Herzfeld,Eyring}, but its fundamental essence is shown above to be ultimately the same: this is the reciprocal of the minimum window within which two successive configurations cannot be distinguished. 
Whenever the favorable spatial configuration reappears, the reaction is possible but not necessarily achieved. It proceeds only when a second condition, energetic, is fulfilled simultaneously. Reciprocally, when the required energy level is reached, no reaction will occur if the configuration is not permissive at the same time. The dimensionless part of the constants is therefore the product of two independent probabilities. The second one, the energetic probability, is the exponential factor of Arrhenius.

\section{The exponential factor is based on the exponential distribution of energy}

\subsection{The exponential distribution of energy links kinetics to statistical mechanics}

The exponent of Arrhenius $ E_{a} $ is generally identified to the difference between the threshold energy $ E^{\ddagger} $ required for the reaction and the energy $ E_{i} $ of the reactant(s) $ i $. This exponential factor can be interpreted as a ratio of probabilities \cite{Michel2018a,Michel2018b}. The probabilistic tool dedicated for connecting rate constants to statistical mechanics is undoubtedly the geometric or exponential distribution which is remarkably sufficient to recover the main results of Maxwell-Boltzmann statistics \cite{Michel2013,Michel2018a}. From a theoretical point of view, the probability ratio described in \cite{Michel2018a} can be interpreted as a conditional exponential probability. In order to explain this, let us start by justifying the use of exponential probabilities in this framework. This probability rigorously defines the maximal randomness, or absence of memory in mechanical systems \cite{Michel2013}. The probability density function (PDF) of the exponential probability is

\begin{equation} \label{expon-pdf} f(\mathcal{E})= \frac{1}{\left \langle \mathcal{E} \right \rangle}\ \large{\textup{e}}^{-\dfrac{\mathcal{E}}{\left \langle \mathcal{E} \right \rangle}} \end{equation} 
where $ \left \langle \mathcal{E} \right \rangle $ is the mean value of the distribution, that is to say in the present case the ratio of the total number of energy units over the total number of particles.  The probability that a randomly picked particle has an energy equal to or higher than a certain threshold $ \mathcal{E}^{\ddagger} $ is obtained by the following integration of the PDF

\begin{equation} \label{Expon} P(\mathcal{E} \geqslant \mathcal{E}^{\ddagger})= \int_{\mathcal{E}=\mathcal{E}^{\ddagger}}^{\infty } f(\mathcal{E}) \ d\mathcal{E} =\large{\textup{e}}^{-\dfrac{\mathcal{E}^{\ddagger}}{\left \langle \mathcal{E} \right \rangle}} \end{equation}  \label{E:gp1}

Another integration of the same PDF allows to determine the fraction of particles with a given number of energy units.

\begin{equation} \begin{split} P(\mathcal{E} = \mathcal{E}^{\ddagger})&= \int_{\mathcal{E}=\mathcal{E}^{\ddagger}}^{\mathcal{E}^{\ddagger}+1} f(\mathcal{E}) \ d\mathcal{E} \\& = \left (1- \large{\textup{e}}^{-\dfrac{1}{\left \langle \mathcal{E} \right \rangle}}\right ) \large{\textup{e}}^{-\dfrac{\mathcal{E}^{\ddagger}}{\left \langle \mathcal{E} \right \rangle}}\\& =\dfrac{\large{\textup{e}}^{-\dfrac{\mathcal{E}^{\ddagger}}{\left \langle \mathcal{E} \right \rangle}}}{\sum \limits_{j=0}^{\infty }\large{\textup{e}}^{-\dfrac{j}{\left \langle \mathcal{E} \right \rangle}}} \end{split}  \end{equation} 

The same result can also be obtained without using the PDF, by a simple subtraction of probabilities as follows
\begin{equation} \begin{split} P(\mathcal{E} = \mathcal{E}^{\ddagger})& =P(\mathcal{E} \geqslant \mathcal{E}^{\ddagger})-P(\mathcal{E} \geqslant \mathcal{E}^{\ddagger}+1)\\
& =\large{\textup{e}}^{-\dfrac{\mathcal{E}^{\ddagger}}{\left \langle \mathcal{E} \right \rangle}}-\large{\textup{e}}^{-\dfrac{\mathcal{E}^{\ddagger}+1}{\left \langle \mathcal{E} \right \rangle}}
\\
& =\large{\textup{e}}^{-\dfrac{\mathcal{E}^{\ddagger}}{\left \langle \mathcal{E} \right \rangle}}\left (1- \large{\textup{e}}^{-\dfrac{1}{\left \langle \mathcal{E} \right \rangle}}\right ) \\ 
&  =\dfrac{\large{\textup{e}}^{-\dfrac{\mathcal{E}^{\ddagger}}{\left \langle \mathcal{E} \right \rangle}}}{\sum \limits_{j=0}^{\infty }\large{\textup{e}}^{-\dfrac{j}{\left \langle \mathcal{E} \right \rangle}}} \end{split} \end{equation}

This result singularly resembles the famous partition function of Boltzmann, which is not necessary for the present model but is shown to recall that the exponential probability is the basis of statistical mechanics in general. Efforts have been made to derive the Arrhenius equation from statistical mechanics \cite{Tolman,Eyring}. The present proposal is an effort of maximal reduction suggesting that statistical mechanics and the structure of the Arrhenius equation have a common origin: the simple exponential probability. The exponential probability is indeed sufficient to rigorously recover the energetic part of the Arrhenius law, as shown below.

\subsection{Bayesian view of the Arrhenius exponential factor}

Consider a chemical mixture containing different types of molecules over which energy spreads according to a unique exponential (random) distribution whose mean value is $ \left \langle E \right \rangle $ per molecule. Each type of molecule $ i $ has a minimal energy content $ E_{i} $ imposed by its specific chemical structure, but it can take higher values $ E \geqslant E_{i} $. In addition, imagine that this molecule can interconvert into a molecule of type $ j $ provided its energy reaches a fixed energy threshold of interconversion $ E^{\ddagger} $. Then, the probability of sufficient energy for molecule transformation is the conditional probability

\begin{equation} P \left(E \geqslant E^{\ddagger} \rvert E \geqslant E_{i} \right) \end{equation}

which is the probability that $ E \geqslant E^{\ddagger} $, given (or knowing that) $ E \geqslant E_{i} $. Obviously if $ E_{i} \geqslant E^{\ddagger} $, this probability is 1, which mans that the reaction is not energetically restricted. But if $ E_{i} < E^{\ddagger} $, then the Bayes theorem gives 

\begin{equation} P \left(E \geqslant E^{\ddagger} \rvert E \geqslant E_{i} \right)=\dfrac{P \left(E \geqslant E_{i} \rvert E \geqslant E^{\ddagger} \right) \  P \left(E \geqslant E^{\ddagger} \right)}{P \left(E \geqslant E_{i}\right)} \end{equation}
and since
$$ P \left(E \geqslant E_{i} \rvert E \geqslant E^{\ddagger} \right) =1 $$
one obtains the probability ratio

\begin{subequations}
\begin{equation} P \left(E \geqslant E^{\ddagger} \rvert E \geqslant E_{i} \right)=\dfrac{P \left(E \geqslant E^{\ddagger} \right)}{P \left(E \geqslant E_{i}\right)} \end{equation}
which can be more directly obtained as the conditional probability

$$\dfrac{P \left(E \geqslant E^{\ddagger} \cap E \geqslant E_{i} \right)}{P \left(E \geqslant E_{i}\right)} $$

and which is, when applied to the exponential probability,
\begin{equation} P \left(E \geqslant E^{\ddagger} \rvert E \geqslant E_{i} \right)=\large{\textup{e}}^{-\dfrac{E^{\ddagger}-E_{i}}{\left \langle E \right \rangle}} \end{equation}
\end{subequations}
This is the traditional form of Arrhenius when assigning to the mean thermal energy $ \left \langle E \right \rangle $ the value $ k_{B}T $. Finally when the required level of energy is reached, the reactant can either fall back to its initial state or fall to the transformed molecule, with identical probabilities 1/2, thus giving the final form of the rate constant

\begin{equation} k= \dfrac{1}{2 \ \Omega \tau} \ \large{\textup{e}}^{-\dfrac{E^{\ddagger}-E_{i}}{\left \langle E \right \rangle}} \end{equation} 

As expected, the energy threshold $ E^{\ddagger} $ disappears from the equilibrium constant.
\begin{equation} \begin{split} K_{ji}&= \dfrac{\Omega_{i}}{\Omega_{j}} \ \dfrac{P(E \geqslant E^{\ddagger})}{P(E \geqslant E_{i})} \dfrac{P(E \geqslant E_{j})}{P(E \geqslant E^{\ddagger})}\\&= \dfrac{\Omega_{i}}{\Omega_{j}} \ \dfrac{P(E \geqslant E_{j})}{P(E \geqslant E_{i})}  \\&= \dfrac{\Omega_{i}}{\Omega_{j}} \ \large{\textup{e}}^{\dfrac{E_{j}-E_{i}}{\left \langle E \right \rangle}} \end{split} \end{equation} 

Substituting $ \left \langle E \right \rangle $ by $ k_{B}T $, we obtain for the waiting time of an elementary event, the following exponential of exponential

\begin{equation} P(X_{i}\geqslant t)= \textup{Exp} \left [-\dfrac{t \ k_{B}T}{2 \ \Omega_{i} \ h} \ \textup{Exp}\left (-\dfrac{E^{\ddagger}-E_{i}}{k_{B}T}  \right )  \right ] \end{equation} 

\subsection{Connection with free energies}

The activation energy of Arrhenius in Eq.(\ref{Arrhenius}) is conceived as the difference between the threshold energy necessary for the reaction to proceed ($ E^{\ddagger} $) and the energy ($ E_{i} $ usually defining the enthalpy in thermochemistry) of the reactant $ i $: $  E_{a}= E^{\ddagger}-E_{i} $. Using this definition, the equation of Arrhenius (Eq.(\ref{Arrhenius})) can be readily identified with the form using free energies, initially written in 1911 \cite{Scheffer}.
\begin{subequations}
\begin{equation} \label{1911}  k = \nu \ \large{\textup{e}}^{-\dfrac{\Delta G^{\ddagger}}{k_{B}T}} \end{equation}
The free energy of activation is

\begin{equation} \label{deltaG}  \Delta G^{\ddagger} = E^{\ddagger}-G_{i} \end{equation}
where $ G_{i} $ is the free energy of the reactant $ i $:
\begin{equation}  G_{i} =  E_{i}-TS_{i} \end{equation}
where $ S_{i} $ is the reactional entropy, which can be connected to the number of configurations of the reactant $ \Omega_{i} $ through the Boltzmann formula for entropy
\begin{equation} \label{Boltzmann} S_{i} =  k_{B} \ln \Omega_{i} \end{equation}
\end{subequations}
Hence, to equalize the formula of Arrhenius Eq.(\ref{Arrhenius}) and Eq.(\ref{1911}) using free energies, the preexponential factor of Arrhenius should be identified as
\begin{equation}   A = \dfrac{\nu}{\Omega_{i}} \end{equation} 

The universal thermal frequency $ \nu $ was then identified to $ \dfrac{k_{B} T}{h} $ after introduction of quantum mechanics in the treatment \cite{Herzfeld,Eyring}, of which a simpler interpretation is suggested in Section 3 (\ref{tau}). The formal equivalence between Eq.(\ref{Arrhenius}) and Eq.(\ref{1911}) is particularly transparent when using the simple entropy of ideal gases. The notion of free energy is essential for predicting the evolution of a whole macroscopic system, but the Arrhenius equation combining entropy and energy into a single entity of free energy is less practical for deciphering its fundamentally probabilistic nature. The separation into distinct notions also makes clearer the fact that when temperature strongly increases, reactions rates become essentially a matter of entropy since the exponential factor tends to unity.

\section{Historical development of the rate constants}

The story of the investigations on the nature of rate constants, culminating with the TST, is remarkably reported in \cite{Laidler1983}. The general form of the exponential factor clearly derives from the pioneering van't Hoff's thermodynamic approach \cite{van't Hoff} and subsequent attempts were aimed at linking it to statistical physics \cite{Tolman,Eyring}.

\subsection{Thermodynamic emergence of the exponential factor}

The current form of the exponential factor has in fact not changed significantly since van't Hoff's thermodynamic approach. van't Hoff initially identified a functional relationship between temperature $ T $ and the equilibrium constant $ K $

\begin{equation}  \dfrac{d \ln K}{dT} = \dfrac{q}{2 T^{2}} \end{equation} 

where $ q $ was defined by van't Hoff as the heat developped by the reaction, which slightly depends on temperature. The heat of a reaction usually refers to a change of enthalpy, but the term $ q/2 $ has then been replaced in textbooks by $ \Delta G^{0}/R $, while still quoting van't Hoff. Since an equilibrium constant is the ratio of forward and backward rate constants $ K_{f} = k_{f}/k_{b} $, 

\begin{subequations}
\begin{equation}  \dfrac{d \ln k_{f}}{dT}-\dfrac{d \ln k_{b}}{dT}  = \dfrac{q}{2T^{2}} \end{equation}
van't Hoff concluded that the rate constants should take a similar form.
\begin{equation} \dfrac{d \ln k_{f}}{dT}  = \dfrac{A_{f}}{T^{2}}+B \end{equation}
\begin{equation} \dfrac{d \ln k_{b}}{dT}  = \dfrac{A_{b}}{T^{2}}+B \end{equation}
with
\begin{equation} \label{deltaA} A_{f}-A_{b}= \dfrac{q}{2} \end{equation}
\label{E:gp4}
\end{subequations}

with a unique value of $ B $ and different values of $ A $ depending on the temperature for the forward and backward reactions. Eq.(\ref{deltaA}) conforms to the fundamental principle of thermodynamics, but it tolerates to insert additional parameters in the rate constants, identical for the reciprocal reactions and eliminated by subtraction. $ A_{f} $ and $ A_{b} $ were then interpreted as activation energies, playing a role of kinetic restriction but without any thermodynamic role. These activation energies were conceived as differences between the energy of molecules actually committed to the reaction and the energy of the reactants before reaction. The former ($ E^{\ddagger} $) is one of the additional parameters mentioned above, identical for the reciprocal reactions.

\subsection{The TST}

Quantum physics has then been decisive in elucidating the value of the elementary frequency of the pre-exponential factor, considered by Eyring himself as his main contribution. In fact, as suggested in \cite{Laidler1983}, the TST is more a compilation of notions scattered in the previous literature than a real novelty and Eyring acknowledged that his formulation has certain features in common with a number of previous treatments \cite{Eyring}. In particular, the identity between the equation of Arrhenius and that derived from the TST is clear when giving to the pre-exponential factor a configurational meaning \cite{Steinfeld}.

\section{Comparison of the parameters used in the different models}

\subsection{Definition of the parameters in the present theory}

Unclear definitions of variables present in equations can result in scientific confusion, particularly in the present context. For instance, the types of energy considered should be specified. Free energies decrease with temperature whereas enthalpies increase with temperature. These opposite behaviors are critical when studying the role of temperature on reaction rates. Hence, to compare the different approaches, the meaning of the different terms should be carefully explicited. In the exponential factor of the present proposal $$ \large{\textup{e}}^{-\dfrac{E^{\ddagger}-E_{i}}{\left \langle E \right \rangle}}  $$ 
\begin{itemize}
\item $ E^{\ddagger} $ is a fixed energy threshold which does not depend on temperature. As such, it cannot be the energy of a molecule, which obliges us to reconsider the conception of the activated complex. In the example of the isomerization of 2-butene, it is the energy required for breaking the $ \pi $ bond, which does not depend on the temperature of the system. A rise in temperature only makes it easier to reach it.
\item $ E_{i} $ is the minimal energy of the reactant molecules $ i $, but not their average energy. No reactant molecule of kind $ i $ can have an energy lower than $ E_{i} $ and this threshold increases with temperature.
\item $ \left \langle E \right \rangle $ is the average single particle energy imposed by the bath to the whole system, including the solvent, which must be specifically recalculated for pure reactants in vacuum.
\end{itemize}

These definitions deny the constancy of the activation energy. Indeed, the exponent of rate constants depends twice on the temperature through $ E_{i} $ and $ \left \langle E \right \rangle $ and that of the equilibrium constants depends three times on the temperature, through $ E_{i} $, $ E_{j} $ and $ \left \langle E \right \rangle $. The access of a molecule to the reactional energy threshold $ E^{\ddagger} $ is favored by an increase in temperature in two ways: (i) by increasing the general average energy, which itself increases the denominator of the exponent, and (ii) by increasing the specific energy of the considered reactant, which decreases the numerator of this exponent. However, the influence of the denominator is clearly predominant for simple reactions with a large $ E_{a} $, for example only dependent on translational energy. In addition, in temperature ranges in which the Dulong-Petit law applies, the numerator remains roughly constant so that the Arrhenius plot approaches a straight line, as often observed in practice. 
\\ It is now of interest to compare the definitions of the different parameters listed above with those of the other theories. 

\subsection{The activation energy}

An accepted interpretation of the activation energy \cite{Tolman,Rice,Menzinger,Truhlar}, is a difference between the average energy of molecules entering into the reaction ($ E_{\textup{activated}} $) and the average energy of the reactants ($ E_{\textup{average}} $), leading to the explicit form of the exponential factor:

\begin{equation} \large{\textup{e}}^{-\dfrac{E_{\textup{activated}}-E_{\textup{average}}}{RT}}  \end{equation}  

This formula is considered valid for monomolecular as well as multimolecular reactions \cite{Tolman}. 

\begin{itemize}
\item $ E_{\textup{activated}} $ is conceived as the mean energy of the molecules undergoing the chemical transformation.
\item $ E_{\textup{average}} $ is the average energy of all the potentially reactant molecules. \\ The view proposed here is radically different.
\end{itemize}

\subsection{The nature of molecular energies in the Arrhenius exponent}

\subsubsection{Molecular energies conceived as minimal, not average.}

Contrary to the previous definition of the energy of the reactants as average energies of the whole assembly of reaction precursors  \cite{Tolman,Rice}, here it is a basal energy. A given molecule $ i $, due to its particular molecular structure and chemical bonds, cannot have an energy lower than a certain level $ E_{i} $ at nonzero temperature. Individual molecules can attain any larger value through thermal fluctuations, with a probability exponentially lower for increasing values. The minimal molecular energy however increases with temperature, just as enthalpies according to Kirchhoff's law. When rationally applied to the exponential distribution, the vision of molecular energy as the average energy would in fact lead to an interesting rate theory whose implications have been exhaustively studied in \cite{Michel2018a}, which may apply to some specific contexts such as two-temperature systems, but does not support a simple test with experimental chemistry \cite{Michel2018b}. In the probabilistic model described here, the average molecular energy is $ \left \langle E \right \rangle $, present in the denominator of the exponent and which does not concern specifically the considered reactant. Its value can be identified to $ k_{B}T $, or $ RT $ when using moles of molecules and replacing Boltzmann contant by the ideal gas constant. The average energy $ \left \langle E \right \rangle $ is an input in the present approach, whereas the identification of $ k_{B}T $ as average energy is a secondary result obtained by calculation in statistical physics, in which this factor is introduced as the inverse of the Lagrangian multiplier $ \beta $. Accordingly, $ k_{B}T $ is omnipresent in kinetics and thermodynamics approaches but not specially described as an average energy, which may explain why the role of average was transferred to the exponent's numerator through $ E_{\textup{average}} $ \cite{Tolman,Menzinger}.
Finally, a more general problem, not specific to rate theories, is that absolute enthalpies can take negative values, contrary to the fundamental definitions of energy. As an alternative, positive energies can be used \cite{Michel2018b}, simply defined with a modified Kirchhoff's law. In solution chemistry at the particular temperature $ T_{1} $, the positive energy of a molecule $ i $ is defined using its temperature-dependent heat capacity $ C_{i}(T) $ as $$ E_{i}(T_{1})= \int_{T=0}^{T_{1}} C_{i}(T) \ dT $$ Indeed, given that translations, rotations and vibrations, disappear in classical physics at zero kelvin, molecular energy is expected to be zero. 

\subsubsection{Metastable equilibrium.}
The recovery of the Arrhenius exponential factor through the conditional probability approach described here, sheds particular light on the nature of molecular energies. From the perspective of reaction paths in a landscape, chemical systems are multistable even in equilibrium and their different components coexist in different wells of stability. In equilibrium, as assumed in the TST, the dominant paths of back and forth reactions are superposable and pass through the saddle points. But out of equilibrium, the global maxima along the dominant paths would no longer be saddle points and would be different for the reciprocal reactions, due to the existence of curls \cite{Wang}. Such a dependence of rate constants on the state of the system would seriously complicate the theory. Moreover, the stability of large molecules in equilibrium is itself questionable. Unlike simple eternal molecules, high-energy molecules have average lifetimes that are not infinite, so that considering them in equilibrium may only be an approximation valid at our time scale. The conditional probability perspective greatly simplifies the situation since molecular energies are just energy pawls over which molecular structures are (even if only temporarily) locked. In this view, complex molecules are exponentially rarer than simple molecules. 

\subsection{The elusive activated complex}
The activated complex is currently envisioned as an energy-rich molecule located on the reaction path between two interconvertible reactants, and the reaction rate is supposed to be proportional to the concentration of this molecule. Eyring explained that this complex corresponds to a saddle point in a potential energy landscape and is like any other molecule, except in the degree of freedom in the negative curvature it is flying. The relatively flat curvatures in the vicinity of the saddle point would allow to identify it with a stable chemical compound \cite{Eyring}. But this intuitive conception obviously ignores the notion of elementary reaction. Let us consider the reversible first order transitions

\begin{center}
\ce{\textit{A}
<=>[\ce{\textit{a}}][\ce{\textit{b}}]
$\ce{\textit{B}}$}
\end{center}

At a given temperature, the rates $ a $ and $ b $ are constants, which are always at work, in transient, steady state as well as equilibrium conditions. The continuous fluxes of molecules from $ A $ to $ B $ and from $ B $ to $ A $, equalize in equilibrium to $$ v = \dfrac{N}{\dfrac{1}{a}+\dfrac{1}{b}} $$ where $ N $ is the total concentration of molecule $ N=[A]+[B] $. Let us now insert an activated chemical complex denoted $ C $ between $ A $ and $ B $. In the TST, $ C $ is supposed in chemical equilibrium with $ A $ for the forward reaction, and it is symmetrically in equilibrium with $ B $ for modeling the backward reaction. Since both reactions occur simultaneously, the so-called "point of no return" assumption to TST is not justifiable and the scheme becomes

\begin{center}
\ce{\textit{A}
<=>[\ce{\textit{a}}][\ce{\textit{d}}]
$\ce{\textit{C}}$
<=>[\ce{\textit{c}}][\ce{\textit{b}}]
$\ce{\textit{B}}$}
\end{center}

In equilibrium, the concentration ratio reads

\begin{subequations}
\begin{equation} 
\dfrac{[B]}{[A]}= \dfrac{a \ c}{b \ d} 
\end{equation} 
but this constant was previously known to be
\begin{equation} 
K_{AB}= \dfrac{a}{b} 
\end{equation} 
\end{subequations} 

which implies that $ d=c $. The scheme then simplifies to

\begin{center}
\ce{\textit{A}
<=>[\ce{\textit{a}}][\ce{\textit{c}}]
$\ce{\textit{C}}$
<=>[\ce{\textit{c}}][\ce{\textit{b}}]
$\ce{\textit{B}}$}
\end{center}

The steady state one-way fluxes are now $$ v= \dfrac{N}{\dfrac{1}{a}+\dfrac{1}{b}+\dfrac{1}{c}} $$ where $ N=[A]+[B]+[C] $. To obtain the accepted result, we must set $ c \sim \infty $. Concretely for the isomerization reaction examined in \cite{Michel2018b}, this intermediate is intuitively assumed to correspond to some highly energized molecule with a broken $ \pi $ bond. But does such a logical intermediate species can be considered as a molecular species? The mere existence of a compound whose lifetime is $ 1/c \sim 0 $ would violate quantum theory. Quantum physics predicts that the successive configurations of a molecular system cannot be distinguished if they are temporally spaced by less than the thermal quantum time step. In the present case no molecule can live less than $ h/k_{B}T $, as recalled in \ref{tau}. The reciprocal of this window is the maximum frequency of all the events, which is more or less slowed down by the entropic and energetic restrictions of chemical reactions, thereby preventing our world from rapidly falling to its state of maximum entropy in a cascade of reactions with a rate of $ k_{B}T/h $.

\subsection{The energy and entropy of the activated complex}
\subsubsection{Energy.}
In the TST, the activation energy is the difference of altitude in the potential energy surface \cite{Eyring} (Fig.1A). The basal axes of this theoretical surface are the popular "reaction coordinates" also called in textbooks "reaction progress", but whose units are not well defined. These coordinates have in particular not the dimension of a spatial length, but the col of the saddle in this landscape is nevertheless crossed by a moving mass point with a velocity defined in the real space (meters/second), thereby mixing theoretical and concrete spaces. Some analogies can however be found between the landscape metaphor and the present theory:

\begin{itemize}
\item $ E^{\ddagger} $ is unique because fluctuations are in theory impossible on a convex ridge contrary to the assumption of Eyring, since they would preclude the return to the initial point.
\item $ E_{i} $ is the minimal energy of the reactant molecules $ i $, because this value is precisely the lowest point at the bottom of an attraction basin. This situation exerts a restoring force in case of deviation. In this respect, this unique molecular energy is not an average but clearly a local minimum. 
\end{itemize}

Things are less clear about the role of temperature, the meaning of $ k_{B}T $ and the capacity of temperature changes to either modify the overall altitude of a rigid landscape or to distort a deformable landscape. In this respect, the letter $ E $ generally used for energy does not explicitly refer to $ H $ or $ G $. Free energy landscapes are subject to more profound reshaping including altitude inversions. For example a reaction such that $ \Delta H <0 $ and $ \Delta S <0 $ is favored ($ \Delta G <0 $) only at low temperature and conversely a reaction such that $ \Delta H >0 $ and $ \Delta S >0 $ is favored ($ \Delta G <0 $) only at high temperature. Moreover, the energy gap between the saddle point and the reactant is called activation energy irrespective of whether the landscape is conceived as a potential of free energy or enthalpy, thereby increasing the confusion. In the dual probabilistic approach presented here, the configurational and energetic probabilities are naturally independent.

\subsubsection{The open question of the transition state entropy.}
In the TST, the transition state has a nonzero entropy, corresponding to a certain number of configurations $ \Omega^{\ddagger} $ sometimes witten $ \Omega_{AB} $, different from unity. The existence or non-existence of this entropy would however not modify the detailed balance laws given that it is systematically eliminated by simplification in equilibrium. The determination of this parameter is notoriously problematic for using the TST. Without definitely deciding on this point, the assumption retained here is that the entropy of a reactional intermediate could be dispensed with, as is implicit in the variation of free energy described in Eq.(\ref{deltaG}) and in accordance with the postulated absence of chemical intermediates. If assuming that the transition state is not a molecule but a single threshold value, it seems logical to postulate that the transition state configuration is also unique in the degree of freedom involved in the reaction, as illustrated in Fig.1B, giving zero reactional entropy and $ G^{\ddagger}=E^{\ddagger} $. The molecular energy $ E_{i} $ used in thermochemistry is the absolute enthalpy ($ H $), which depends on temperature, but as this is not the case for the transition state, it could be preferable to assign to the transition state a symbol like $ E^{\ddagger} $ instead of $ H^{\ddagger} $. The degrees of freedom with orthogonal components are expected to contribute to reaction energy according to the equipartition principle, but their participation to reaction entropy is more questionable. If a configuration that is not parallel to the dominant reaction coordinate, allows nevertheless the reaction to proceed, the global maximum energy of this path is expected to be higher than the saddle point threshold.

\subsection{Maintenance of energy distribution in and out of equilibrium}
One of the objections to the TST in the early stages was the equilibrium between the activated complex and the reactant \cite{Laidler1983}. The probabilistic treatment used here cancels this problem since the chemical equilibrium of Eyring is replaced by a fluctuation of energy, that is not subjected to the maximum frequency of chemical reactions. The Boltzmann distribution is preserved out of equilibrium, so that rate constants remain constant in equilibrium, transient and steady state conditions.

\begin{center}
\includegraphics[width=8.6cm]{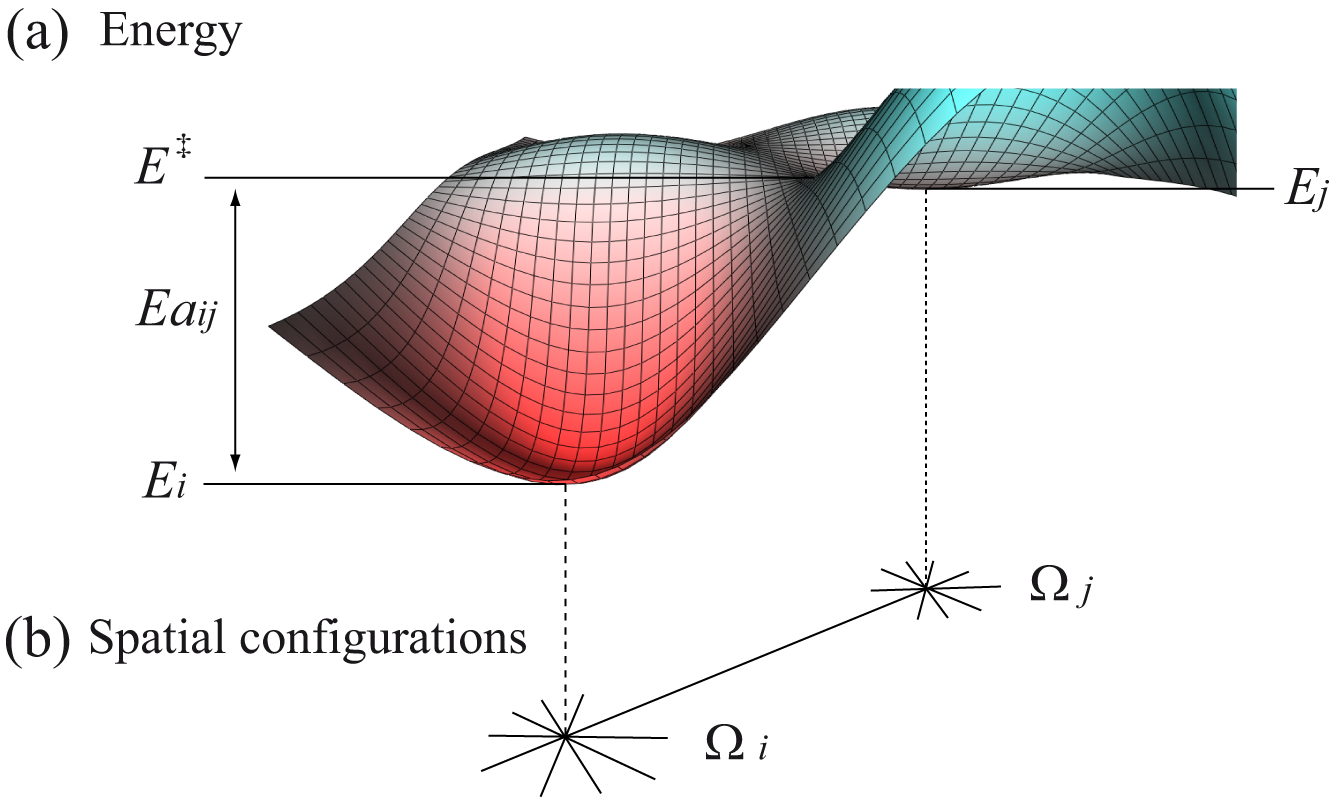}
\end{center}
\begin{small} \textbf{Figure 1}. Dissociated views of energy and configurations in the TST. \textbf{(a)} The energy level of the saddle point in the potential energy landscape, should be reached to allow the reaction. Temperature rise is expected to reduce the activation energy and accelerate the reactions, in part by increasing the basal reactant energies $ E_{i} $ and $ E_{j} $ while $ E^{\ddagger} $ remains constant. \textbf{(b)} The interconvertible reactants have multiple spatial configurations symbolised by the straight lines representing molecular orientations, out of which only one allows the reaction to proceed. The unique reactional configuration connects the two interconvertible reactants, so that it is expected to be also that of the transition state, in line with an absence of transitional entropy. \end{small}\\

\section{Discussion}

The presentation of the familiar form of rate constants proposed here from a simple and robust probabilistic framework, has the pedagogical advantage to not recourse to questionable assumptions of the TST such as (i) the definition of the activated complex as a chemical intermediate in equilibrium with the reactant and (ii) the so-called "point of no return" hypothesis which is  poorly justifiable theoretically. In addition, the alternative view proposed here makes unambiguous the definitions of the parameters influencing kinetics, including activation energy and molecular energy, because they are entering quantities in the present approach, whereas they were interpretations in former rate theories. Imprecise definitions in the field of rate theories increase confusion for both students and researchers \cite{Pacey} and make difficult to rigorously test the models. Historically, once the Arrhenius formula was taken for granted owing to its capacity to satisfactorily describe chemical kinetics, the aim of most studies was to justify the formula by interpreting its ambiguous terms, including activation energy initially defined as the slope of the Arrhenius plot \cite{Truhlar,Pacey}. By contrast, these components are starting ingredients of the reconstruction of the Arrhenius formula proposed here, and as such are rigorously defined and cannot give rise to varied interpretations. This de novo probabilistic view bypasses complex notions of rate theories, which were in fact not devoid of a good deal of empiricism \cite{Laidler1983}. Another example of internal contradiction is that the calculation of motion along reaction coordinate makes use of temporal parameters, although the resulting rate constants found in this way are precisely supposed to introduce time in systems. \\
Contrary to the historical equations of Arrhenius and van't Hoff, the new relations described here between energy, kinetic and equilibrium constants are based on Boltzmann's laws of energy distribution, supposedly random and thus necessarily describable using the exponential probability. This re-enactment of the Arrhenius equation draws direct theoretical and pedagogical links between kinetic rates and statistical distributions. Rate constants are conditional probabilities rooted in Maxwell-Boltzmann statistics, which can itself be reduced to exponential probabilities.\\

\end{multicols}{2}
\end{document}